# Janus icosahedral particles: amorphization driven by three-dimensional atomic misfit and edge dislocation compensation

Zhen Sun[1,3], Yao Zhang[1,3], Zezhou Li[1], Xuanxuan Du[1], Zhiheng Xie[1], Yiheng Dai[1], Colin Ophus[2], Jihan Zhou[1]

[1]Beijing National Laboratory for Molecular Sciences, Center for Integrated Spectroscopy, College of Chemistry and Molecular Engineering, Peking University; Beijing, 100871, China.

[2]National Center for Electron Microscopy, Molecular Foundry, Lawrence Berkeley National Laboratory, Berkeley, CA 94720, USA.

[3]These authors contributed equally to this work.

Correspondence and requests for materials should be addressed to J. Z. (email: jhzhou@pku.edu.cn)

**Icosahedral nanoparticles composed of fivefold twinned tetrahedra have broad applications. The strain relief mechanism and angular deficiency in icosahedral multiply twinned particles are poorly understood in three dimensions. Here, we resolved the three-dimensional atomic structures of Janus icosahedral nanoparticles using atomic resolution electron tomography. A geometrically fivefold face consistently corresponds to a less ordered face like two hemispheres. We quantify rich structural variety of icosahedra including bond orientation order, bond length, strain tensor; and packing efficiency, atom number, solid angle of each tetrahedron. These structural**

**characteristics exhibit two-sided distribution. Edge dislocations near the axial atoms and small disordered domains fill the angular deficiency. Our findings provide new insights how the fivefold symmetry can be compensated and the geometrically-necessary internal strains relived in multiply twinned particles.**

Introduction

Twinning of tetrahedra to form decahedra or icosahedra is common in multiply twinned particles (MTPs). These structures have broad applications in catalysis, optics, electronics, and electrochemistry[1–5]. The final crystallographic symmetry is determined by a subtle balance between surface and volume contributions to the total energy of the MTPs[6,7]. The associated lattice misfit strain in decahedra or icosahedra has attracted particular interest because of its crystallographically forbidden fivefold symmetry[8–12]. To fill the concomitant space gaps between two adjacent (111) faces of the face-centered cubic (fcc) regular tetrahedron with an angle of 70.53°, internal distortion is incorporated to form decahedral or icosahedral MTPs[11,13–15]. Strain relief leads to disclination and fivefold twins[16,17], accompanied by surface defects like groove structures[18,19] and internal defects such as dislocations[20]. Despite numerous experimental and computational studies on the structure and growth pathways of decahedra and icosahedra[11,21–24], the three-dimensional (3D) atomic structures and misfit strain of icosahedral MTPs remain long-standing problems.

The elastic strain energy resulting from the atomic distortion can be compensated by reducing surface energy through optimal atomic arrangements or dislocations and disclinations[6,7]. Ino and Marks proposed extra crystal planes on the classical decahedron models to reduce surface energy. They achieved this either by exposing more (100) faces[25] or introducing re-entrant planes on the twin boundaries[26]. Both models have been recently confirmed in core-shell decahedral MTPs, where extra high-index crystal planes at the corners or edges further reduce the surface energy[27]. During the growth of icosahedra, faulted islands and grooves have been observed[18,19,28]. These structures form curved surfaces that lower the surface energy and relieve strain[7]. Dislocations and disclinations are observed phenomena in MTPs[12,16,29–31]. Electron tomography offers a method to image dislocations[32] and visualize successive twinning in 3D[33]. However, the atomic misfit, the associated strain relief mechanism, and the bridging solid-angles in icosahedral MTPs are not yet quantitatively understood in 3D[34–36], primarily due to a lack of experimentally obtained 3D atomic structures of MTPs. Here, we use gold-palladium icosahedral nanoparticles (ICNPs) as a model and employ atomic resolution electron tomography (AET)[37–41] to determine the 3D atomic coordinates and atomic packing of icosahedra-like MTPs.

Results

**Janus ICNPs with $I_h$ center atom**

Gold MTPs, coated with a thin shell of palladium, were synthesized[42] and deposited on thin $Si_3N_4$ film for high resolution imaging (Methods). Scanning transmission electron microscopy (STEM) images (Supplementary Fig. 1) reveal that the MTPs exhibit various features, including fivefold twinning boundaries, stacking faults, and randomly-oriented polycrystalline grains. We obtained the atomic coordinates of three ICNPs using AET with the following procedure. The tomographic tilt series of the particles (Supplementary Figs. 2-4) were acquired with an aberration-corrected scanning transmission electron microscope in annular dark-field mode. After drift correction and denoising, the tilt series were reconstructed; and the 3D atomic coordinates and chemical types of all atoms were traced and classified (Supplementary Table 1, Methods). Rather than forming a geometrically perfect icosahedron with six fivefold axes, each of which goes through the $I_h$ center, the ICNPs exhibit two hemispheres with distinct Janus faces. We focus on ICNP-1, which has one hemisphere featuring six ideal fivefold axes labeled as C5. The opposite hemisphere exhibits six pseudo-fivefold axes labeled as C5'. As illustrated by the bond orientation order (BOO) parameter (Fig. 1a and b), one face has higher order than the other. The C5 side exhibits a geometrically nearly-perfect icosahedral face with ten tetrahedra (Fig. 1c and Supplementary Fig. 5a). The C5' side contains stacking faults and edge dislocations at the twin boundaries, achieved through grain boundary slipping and edge dislocation insertion (Fig. 1d, Supplementary Fig. 5b, Supplementary Fig. 6, and Supplementary Movie 1). The adjoining tetrahedral domains share a hexagonal closest packed (hcp) twin boundary, and five domains share a common edge (Fig. 1c and d, and Supplementary Fig. 5). Out of twenty tetrahedra in ICNP-1, eighteen have fcc single crystal structures with varied size (Supplementary Table 2), except for two small domains on the C5' side. The normalized local BOO parameters of the atoms are around 0.5 within the two domains (Fig. 1d, and Supplementary Fig. 7a and b), indicating their amorphous nature (Methods). The pair distribution function (PDF) of each domains indexed from 1 to 20 exhibits the transition from crystalline fcc structure to amorphous structure (Supplementary Fig. 8a).

In ICNP-1, the twelve axes—comprising six C5 and six C5' axes—converge precisely at one center atom, forming the intersection (Fig. 1e). The center atom, exhibiting $I_h$ symmetry, is shared by two intersecting decahedra. The twelve coordinated $D_h$ atoms collectively form a distorted icosahedron (Fig. 1f). The C5' axes are noticeably curved, bending away from the central direction, contrasting with the straight extension of the C5 axes (Fig. 1g, Supplementary Fig. 9, and Supplementary Movie 2). An inherent spatial gap of 7.35° in every fivefold direction needs to be filled when packing fcc tetrahedra to form the icosahedron[36]. We observed three types of distortions near the axes to fill the spatial gap. These include homogeneous expansion (Fig. 1h), pure edge dislocations (Fig. 1i), and edge dislocation accompanied with one-atom-distance shift of the fivefold axes (Fig. 1j). The six C5 axes (Fig. 1h) share a similar expansion pattern, characterized by a local twelve-coordination environment, forming two decahedra like the fivefold axes in decahedral nanoparticles[27]. Conversely, the particle's other side features edge dislocations, which expand the angles of fivefold axes to form pseudo-fivefold C5' axes (Supplementary Fig. 6). These C5' axes exhibit a deviation from the regular decahedral configuration, forming a distorted coordination polyhedron with varied coordination numbers (Fig. 1i and j). Additionally, the edge dislocation squeezes into the neighboring crystal domains, resulting in a significant increase in the distance between two atoms in the same layer (connected with dashed red line in Fig. 1i and j); these two atoms are no longer bonded (Fig. 1i and j). In three of the six C5' axes (axis-1,3,6 in Supplementary Fig. 9), an extra atomic column inserts into the decahedra, resulting in thirteen-coordination environment (Fig. 1i). The remaining three C5' axes demonstrate further distortion (axis-2,4,5 in Supplementary Fig. 9), where the edge dislocation disrupts the original axes, forming a new parallel fivefold atomic column comprised of $D_h$ atoms (Fig. 1j). The hcp atom in the original C5' axes direction shares a vertex (as marked in Fig. 1j) with the $D_h$ atoms located at the parallel column of C5' axis (Fig. 1j). Similar Janus structure is observed in ICNP-2, possessing twelve axes and a geometrically perfect $I_h$ center (Supplementary Fig. 10). In this type of ICNPs, it is interesting that a simple expansion on the C5 side is invariably paired with a complex C5' side containing edge dislocations.

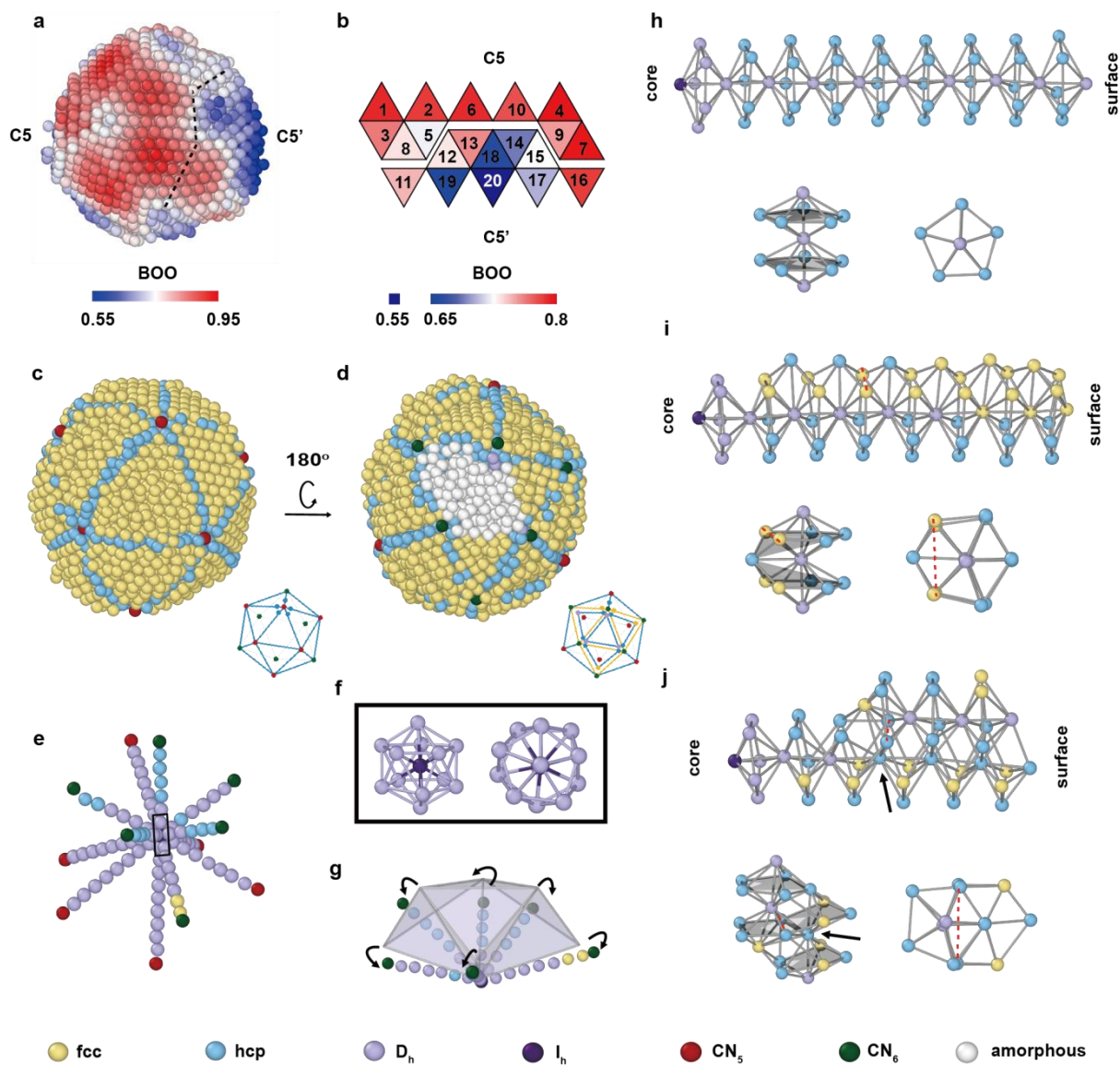

**Fig. 1 | 3D atomic structures and fivefold axes of ICNP-1.** (**a**) Distribution of the BOO parameter in the Janus particle, showing C5 and C5' faces. (**b**) Unwrapped surfaces of twenty domains (each tetrahedral domain has been assigned a number) of ICNP-1 with averaged normalized BOO parameters separated from the C5 and C5' side. (**c, d**) 3D atomic structure of ICNP-1 viewed from two distinct angles. Local atomic coordination environments are colored by the legend at the bottom. Insets show perspective views of the grain boundary frame, revealing fivefold and sixfold ending atoms. (**e**) Twelve axes of ICNP-1, consisting of

six C5 and six C5' axes, with ending atoms marked as red and green, respectively. (**f**) Front and top views of the center icosahedron, enlarged from the black box in (**e**). (**g**) Illustration of the curved C5' axes, bending away from the original straight directions where C5 axes extend. (**h** to **j**) The coordination environment of axial atoms of C5 (**h**) and two types of C5' axes (**i** and **j**). Front and top views of the repeating coordination units are below the column of atoms, respectively. The grey shadows highlight the coordination planes above and below the center atom.

**Two-sided distribution of structural characteristics**

To quantitatively compare the differences between the two faces of the Janus ICNP-1, we analyze the distortion in the twelve axes and solid angle of each crystal domain. We quantified the deviations in three specific bond lengths within the decahedron in twelve axes relative to the standard Au-Au bond length. These bonds include the capping atom bond α, the capping-ring atom bond β and the ring atom bond γ (Fig. 2a). The deviations of all three bond lengths are less pronounced in the C5 axes compared to the C5' axes. The deviation maps show that the capping atom bond α is compressed while the ring atom bond γ is stretched to fill the gap on the C5' side (Fig. 2b, and Supplementary Fig. 11). The crystal lattice on the surface of nanoparticles is more compressed in the radial direction due to the decreased coordination and surface tension[7,27]. The distance between two neighboring $D_h$ atoms, α, is more and more compressed (Fig. 2b) from C5 axes to C5' axes. β remains compressed in C5 axes but becomes stretched in C5' axes due to the bending of the atom columns of C5' axes (Fig. 1g). The γ bonds are stretched in both C5 and C5' axes. The stretching of γ is particularly severe due to the insertion of edge dislocations in C5' axes, with 9% of increasing from the center to the surface of the C5' axes (Fig. 2b). The expansion of γ fills up the angular and spatial gap in the axes. Janus ICNP-2 has similar two-sided distributions in bond length (Supplementary Fig. 12a-e).

To investigate the space-filling mechanism in icosahedra-like MTPs, we determined the solid angles of all twenty tetrahedra within ICNP-1 (Fig. 2c-e, Methods). We assigned numbers to all the twenty tetrahedra and connected adjacent ones, constructing a dodecahedral framework (Fig. 2d). Domains 1 and 20, located at the central positions on the C5 and C5' sides, respectively, are symmetrically distributed within ICNP-1. On the C5 side, the solid angles of most crystal domains are close to 31.6° which corresponds to the angle in the standard fcc lattice. However, the solid angles of domains on the C5' side are predominantly larger than 36°, corresponding to the angle in geometrically perfect icosahedron (Fig. 2c). It's notable that the solid angles follow a hierarchical distribution (Fig. 2e, and Supplementary Table 4). The central domain 20 has the largest value of 50.6° while its three adjacent domains (domains 17, 18, and 19) are in the group with the second largest solid angles (39.6°, 40.5° and 39.6°, respectively). The atom numbers vs. solid angles of all twenty tetrahedra show different trends for C5 and C5' sides. The tetrahedra on the C5 side pack closely with the standard fcc structure, containing more atoms, whereas those on the C5' side have fewer atoms but exhibit larger solid angles (Fig. 2f). Rather than uniformly expanding to fill the inherent angular gap, several tetrahedral domains in ICNP-1 adopt solid angles larger than 36º to compensate for the 3D deficiency. We calculated the atomic packing efficiency (PE) of all tetrahedra in ICNP-1. Although all of them are smaller than the PE of perfect fcc packing, the averaged PE show a two-sided distribution and drops about 5% from C5 side to C5' side (Supplementary Table 6). The PEs of domains 19 and 20 are lower than random close packing (64%)[43,44], also indicating the atoms are loosely packed in these two domains to form amorphous structures. ICNP-2 has similar two-sided distribution in solid angles and PE too (Supplementary Fig. 12f, Supplementary Tables 5 and 7).

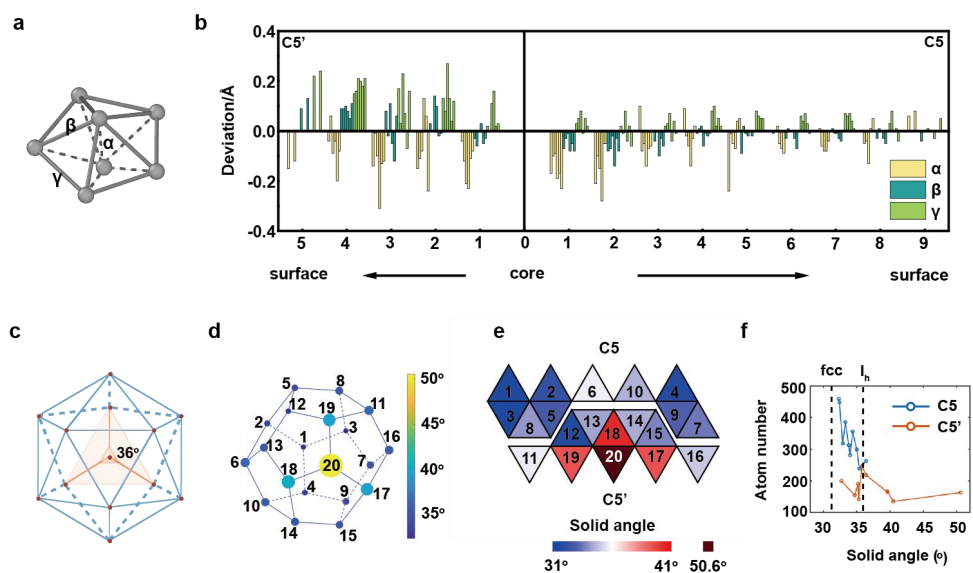

**Fig. 2 | The distinct structural characteristics distinguishing the C5 side and the C5' side within ICNP-1.** (a) An ideal decahedron consisting of three types of atomic bonds. The α, β and γ represent the capping, capping-ring and ring atom bonds, respectively. (b) The deviations of averaged bond lengths of α, β, and γ from the outer surface of C5' axes to the outer surface of C5 axes, by subtracting the standard Au-Au bond length (2.88 Å). (c) The schematic of the solid angle of an ideal tetrahedron in a regular icosahedron. (d) The solid

angle distribution of twenty crystal domains in ICNP-1. The dodecahedral framework is constructed by connecting adjacent tetrahedra. The color and size of the vertices represent the magnitude of solid angles. (**e**) Unwrapped surfaces of twenty domains of ICNP-1 with of solid angles in both C5 and C5' sides. (**f**) Atom number of each domain is plotted against the solid angle in both C5 and C5' sides.

**Strain tensor distributions in C5 and C5' side**

Strain tensor maps were measured based on the 3D coordinates of ICNPs[45], and local strains between the C5 and C5' sides of the whole icosahedron were compared. All six components of the full strain tensor in ICNP-1 exhibit block-like distribution, each corresponding to an individual tetrahedron (Fig. 3a and b). Generally, the strain magnitudes are larger on the C5' side. The principal strain $\varepsilon_{xx}$ is mixed compressive and tensile while $\varepsilon_{yy}$ is almost compressed throughout the particle. Additionally, the distributions of shear strains $\varepsilon_{yz}$ and $\varepsilon_{xz}$ on the C5' side exhibit a bimodal pattern (Fig. 3c), indicative of the expansion of the small crystal domains. Radial strain distributions in ICNP-1 reveal increasing and more scattered strain from the core to the surface (Fig. 3d), suggesting a gradual stress increase within the particle to preserve geometric configuration during growth. Beyond edge dislocations, distortion of fcc lattice on the C5' side crucially addresses angular mismatch during fcc tetrahedra packing into an icosahedron. While the overall strain configuration and dislocations present in ICNP-1 can account for most of the concomitant space gaps, two tetrahedra on the C5' side become amorphous to bridge the remaining space gap. ICNP-2 exhibits similar strain tensor distributions (Supplementary Fig. 14). Coincidentally, amorphization of two tetrahedral domains is observed in ICNP-2 (Supplementary Fig. 7c and d, and Supplementary Fig. 10).

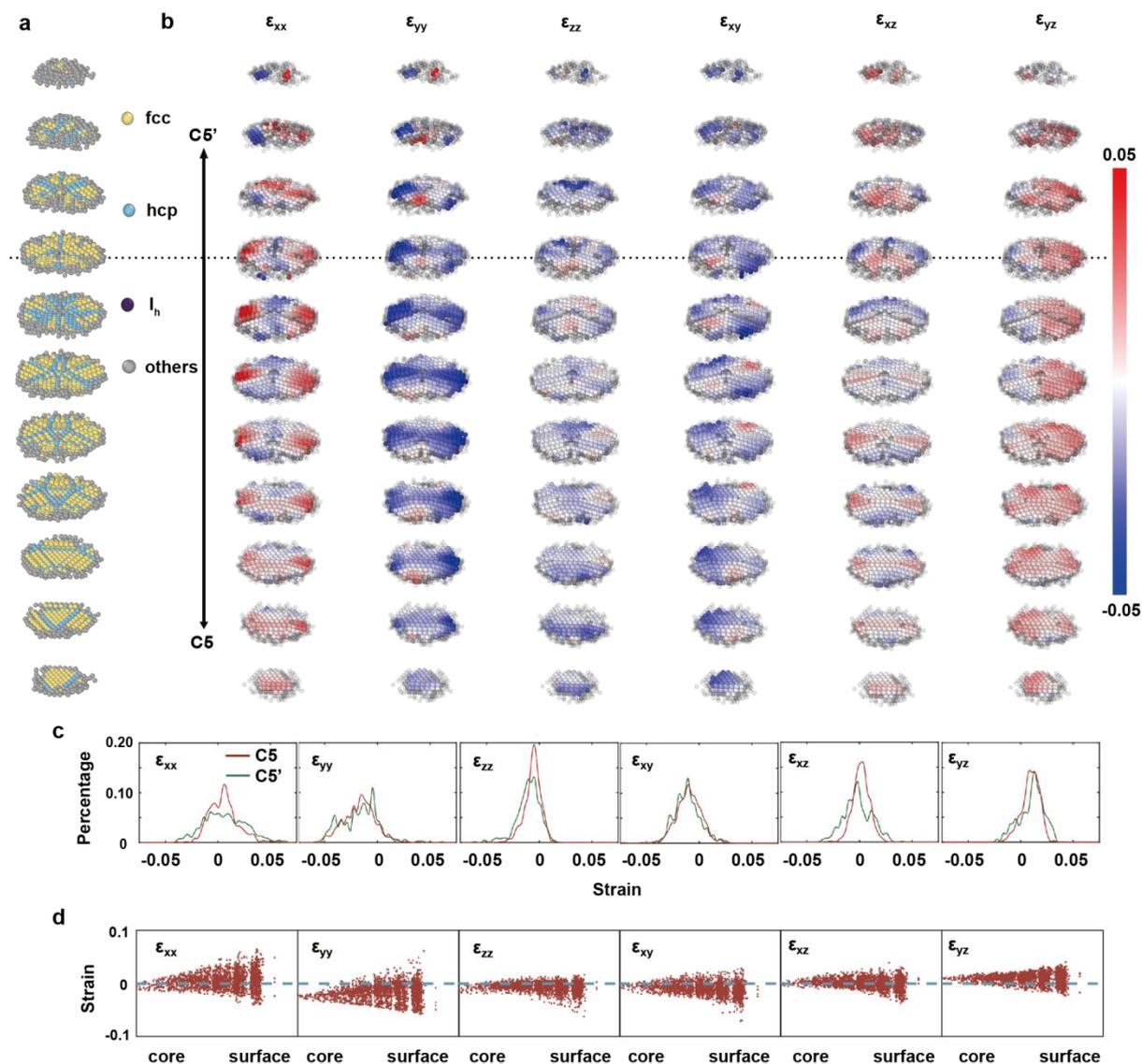

**Fig. 3 | The full strain tensor distributions of ICNP-1.** (a) Atoms in ICNP-1 used to determine the 3D strain tensor, where the atoms in grey from the amorphous domains and the

surfaces are excluded for strain measurement. The Z-axis is determined parallel to the perpendicular direction of the {111} planes in domain 1 and domain 20 (Supplementary Fig. 13). (**b**) Maps of the six components of the full strain tensor, with the same block-like distribution as crystal domains in (**a**). The grey atoms show where the strain tensor cannot be determined due to the lack of reference lattice. (**c**) The histogram of six components of the full strain tensor on both C5 and C5' sides, the strain tensor is mostly larger on the C5' side. (**d**) Scatter plot of six components of the full strain tensor vs. distance from core to surface, with gradual increase and more scattered distribution.

**Janus ICNPs without $I_h$ center atom**

We discovered another type of Janus ICNPs without $I_h$ center atom (Fig. 4a and b). ICNP-3 is a Janus particle with eight axes; these includes three fivefold (C5) axes, three pseudo-fivefold (C5') axes and two sets of twin axes (Supplementary Fig. 15). Notably, these axes do not converge into a single common $I_h$ atom. In ICNP-3, we designate the two faces, A and B. Unlike ICNP-1 and ICNP-2, ICNP-3 lacks symmetrically distributed C5 side and C5' side. Face A exhibits an icosahedra-like structure, composed of ten fcc tetrahedra with three C5 axes and two sets of twin axes (Fig. 4a), contrasting with the six C5 axes in ICNP-1 and ICNP-2. Face B consists of four distorted tetrahedra and three large fcc domains; each large domain is made of two tetrahedra-like grains without hcp grain boundaries (Fig. 4b). Three small tetrahedra-like grains have an amorphous structure (Fig. 4b, and Supplementary Fig. 7e and f). The hcp grain boundaries, combined with specific fcc atoms coordinated to the C5' axes, construct an icosahedra-like framework (Supplementary Fig. 15c). Instead of one-atom-distance, the hcp grain boundaries slip two-atom-distance in this particle, breaking the fivefold axes to form a morphology of splitting "3 hcp + 2 hcp" grain boundaries around the twin axes (Supplementary Fig. 15a and b). In C5 +C5' twin axes, one column composed of $D_h$ atoms and the other column composed of pseudo-$D_h$ atoms are split by two columns of fcc atoms (Fig. 4c and e, and Supplementary Fig. 15a). The twin axes of C5'+C5' is composed of two columns of pseudo-$D_h$ atoms. One column is twelve-coordinated with 7 atoms forming a top ring and 5 atoms forming a bottom ring (left blue boxes in Fig. 4f); the other column possesses the same thirteen-coordinated environment (right orange box in Fig. 4f) as the cluster in Fig. 1i. In addition to a group of splitting "3 hcp + 2 hcp" grain boundaries, two hcp atomic layers also slip by one-atom-distance (Supplementary Fig. 15b). The twin axes end in a disordered boundary domain composed of $D_h$ atoms and hcp atoms (Supplementary Fig. 15e) and then continue to connect with a large grain in which most atoms have an fcc structure. The disordered boundaries are formed of fivefold and sixfold skeletons, corresponding to the orientation of axes (Supplementary Fig. 15d), marked as red and green respectively (Fig. 4c and d, and Supplementary Fig. 15e and f). Comparing to ICNP-1&2, ICNP-3 has a less ordered side with completely different morphology, suggesting multiple pathways can occur during the growth of icosahedra-like MTPs.

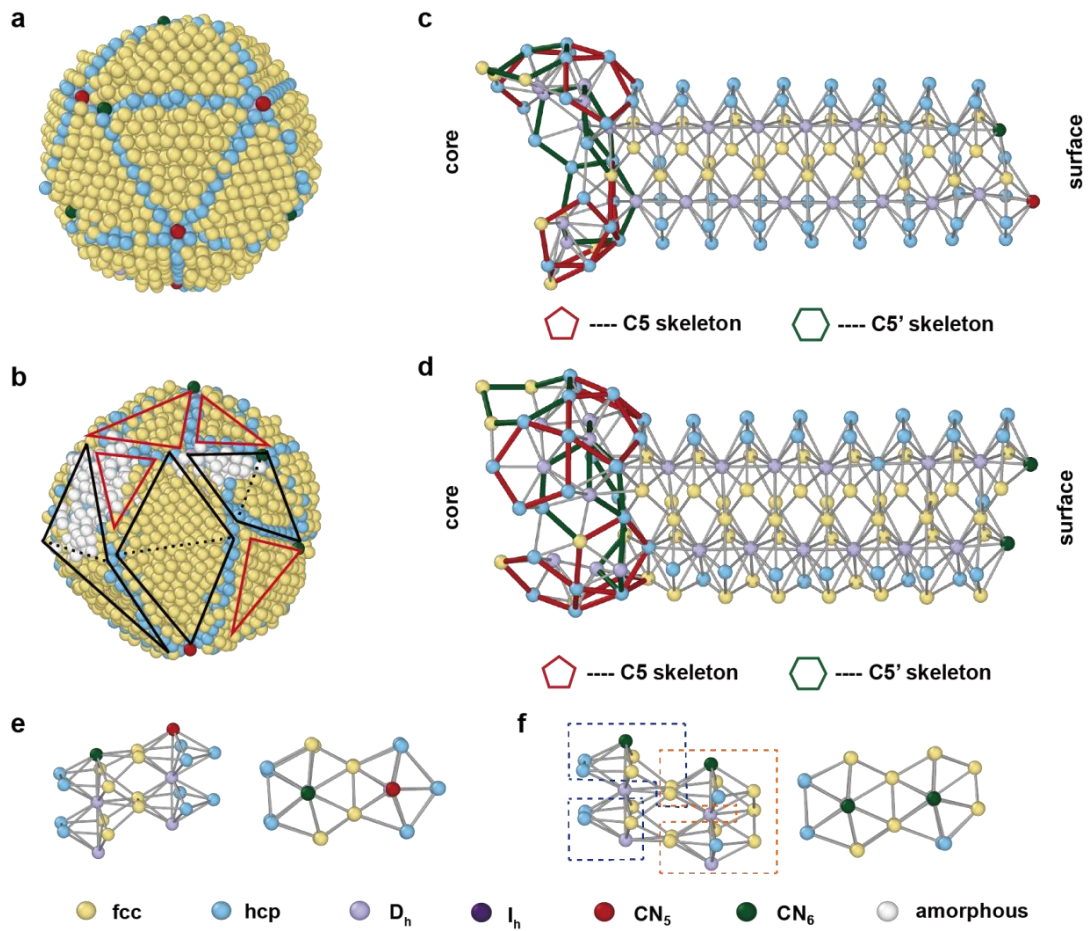

**Fig. 4 | 3D atomic structure of the ICNP-3 without $I_h$ center atom. (a, b)** The two distinct faces of the icosahedra-like particle with eight axes, including three fivefold axes, three pseudo-fivefold axes and two sets of twin axes. In Face B, each of four distorted tetrahedra is marked with a small red triangle. Three large fcc domains are marked with black lines; the black dash lines illustrate the two tetrahedra-like grains without clear hcp grain boundaries. **(c, d)** The atomic structures of two sets of twin axes, C5+C5' **(c)** and C5'+ C5' **(d)**. The bonds in five atom ring and six atom ring are marked as red and green, respectively. **(e, f)** The front view and top view of the repeated coordination unit of **c** and **d**, respectively. The dashed line box in **f** highlight the coordinated atoms of the center pseudo-$D_h$ atom.

**Liquid-solid phase transition of gold nanoparticles**

Our observations suggest there are at least two types of icosahedra-like MTPs with Janus morphology: one possesses a geometrically perfect structure and with an $I_h$ center atom, as in ICNPs 1&2, while the other lacks an $I_h$ center atom, as in ICNP-3. To corroborate our experimental observation, we performed molecular dynamics (MD) simulations on the liquid-solid phase transition of gold nanoparticles using the large-scale atomic/molecular massively parallel simulator (LAMMPS). By quenching gold nanoparticles with similar size to ICNP-1 from 1500 K to 300 K, 100 times, we obtained 100 different structural configurations, comprising four major types: icosahedra (IH, 61%), decahedra (DH, 8%), crystals with stacking fault (SF, 19%) and polycrystals (PC, random MTPs, 12%) (Fig. 5a-c, and Supplementary Fig. 16). We find that the majority of the final structures (61%) exhibit icosahedra-like, Janus morphology with two distinct faces. A geometrically more icosahedra-like hemisphere consistently contrasts with a corresponding hemisphere that displays disordered morphology (Fig. 5b). The potential energy of IH configurations is comparable to that of PC configurations. However, it is larger than the energy observed in both DH and SF configurations. We have compared the averaged BOO parameter of all atoms with the potential energy of all 100 configurations, finding that the more fcc-ordered particles possess the lower potential energy (Fig. 5d). This observation indicates that the IH structures obtained in MD simulations corroborate with our experimental ICNPs; the particle configurations fluctuate with the annealing conditions, and that the IH conformation is governed by atomic diffusion kinetics[46].

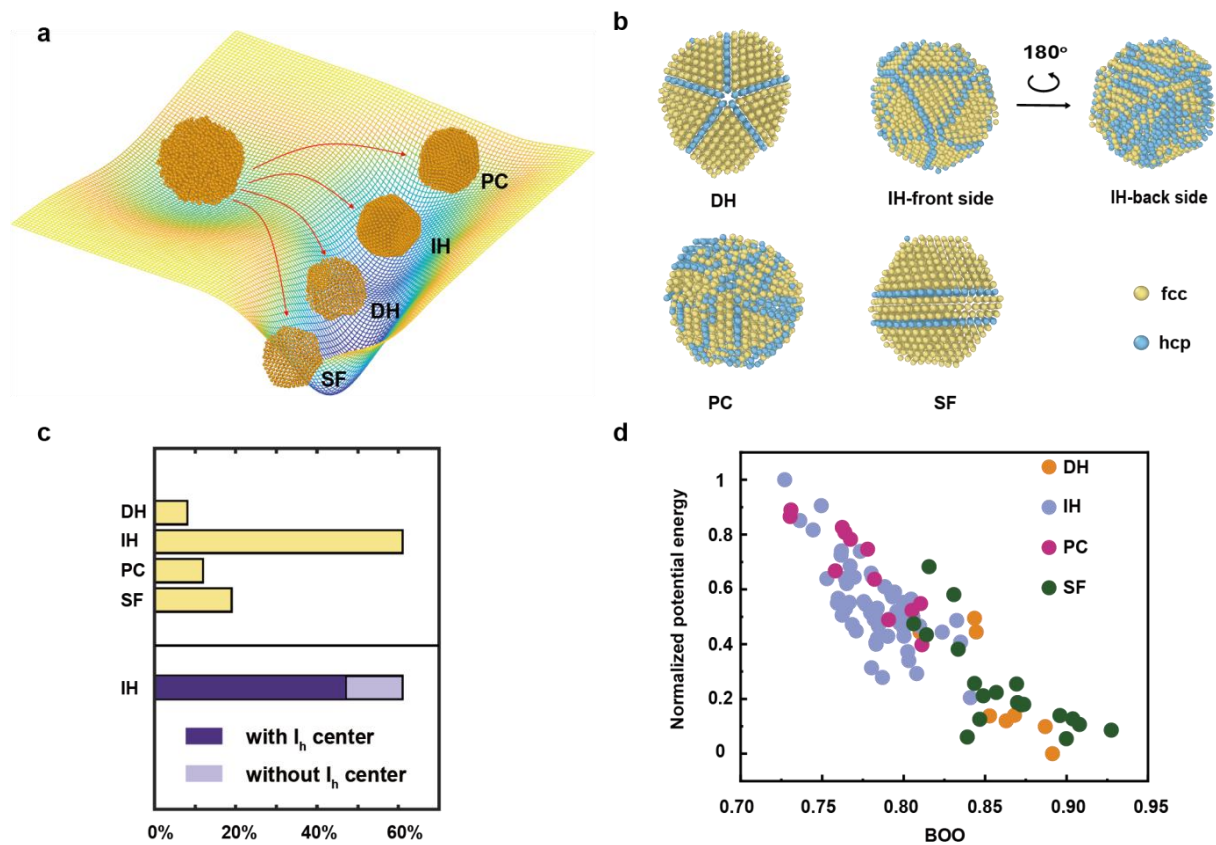

**Fig. 5 | MD simulation of liquid-solid transition of gold nanoparticles.** (**a**) The schematic of the quenching process of gold nanoparticles from 1500K to 300K. (**b**) The four major types of morphologies in annealed gold nanoparticles with 3580 atoms. The icosahedra-like particle has two distinct faces like experimental ICNPs. (**c**) The histogram of each type of morphology and the proportion of the configuration of IH particles with or without $I_h$ center atom. (**d**) The normalized potential energy plots against the averaged BOO parameter of each simulated particle.

**Discussion and Conclusion**

Packing of fcc tetrahedra can form an icosahedron with Janus morphology, exhibiting two distinct faces. Our findings reveal two mechanisms that compensate for the inherent atomic misfit and angular deficiency in icosahedra-like MTPs. These are: i) Inserting an edge dislocation in the C5' axes to alter the original axial atomic coordination. The edge dislocations cause the axial atoms either to become thirteen-coordinated pseudo-$D_h$ atoms, or to slip into the adjacent parallel column to form a new axis. Additionally, the edge dislocations compress the C5' axial atoms, causing them to curve and bend away from their original axial direction. This deformation induces shear strains $\varepsilon_{xz}$ and $\varepsilon_{yz}$ with different mean values between the C5 and C5' side of the ICNPs (Fig. 3c). ii) Sacrificing several tetrahedra-like crystal domains to become amorphous, thereby releasing the strain. The grains enclosed by the C5' axes are smaller and more easily to collapse into a disordered amorphous structure when adjacent edge dislocations are inserted to release large internal strain. The disordered amorphous domain with the largest solid angle is 14.6° larger than the ideal fcc domain, relaxing a large amount of strain and filling the largest angular defects.

In conclusion, we resolved the 3D structures of Janus icosahedral MTPs using AET. A geometrically near-ideal fivefold face is consistently paired with a less ordered face, forming two hemispheres in the ICNPs. Edge dislocations near the axial atoms and small disordered domains with much lower packing efficiency fill the angular deficiency in the icosahedron. Edge dislocations alter the coordination of the axial atoms, causing the bond lengths of the coordination polyhedra to change accordingly to accommodate the angular expansion. Our strain analysis indicates that the internal strain on the C5' side is significant enough to introduce amorphization in one or more of the tetrahedra-like domains. The amorphization of certain small crystalline domains plays a crucial role in strain relaxation and angle filling. Our deep structural analysis of icosahedra-like MTPs provide new insights into how the fivefold symmetry can be compensated and the geometrically-necessary internal strains relived in MTPs.

**Acknowledgments** We thank the support of High-performance Computing Platform of Peking University and the Electron Microscopy Laboratory at Peking University for the use of the aberration-corrected electron microscope. This work was supported by the National Natural Science Foundation of China (Grant No. 22172003). Work at the Molecular Foundry was supported by the Office of Science, Office of Basic Energy Sciences, of the U.S. Department of Energy under Contract No. DE-AC02-05CH11231.

**Author contributions** J. Z. conceived the idea and directed the study. Z. L. and Z. X. acquired the tomographic tilt series. Z. S. performed the imaging processing and reconstructions, and atom tracing. Y. Z. conducted MD simulations. Z. S. and Y. Z. conducted data analysis. Z. S. and X. D. synthesized Au NPs. Z. X. and X. D.

assisted with imaging reconstructions. Z. L., Y. D. and C. O. assisted with data analysis. Z. S., Y. Z. and J. Z. wrote the manuscript. All authors commented on the manuscript.

**Competing interests** The authors declare no competing interests.

# Methods

**Sample preparation**

The gold seeds are synthesized followed by literature[42]. 82.3 mg (0.2 mmol) $HAuCl_4 \cdot 3H_2O$ was dissolved in 7 ml cyclohexane and 7 ml oleylamine, stirred for 15 min at room temperature. 34.8 mg (0.4 mmol) TBAB was dissolved in 1 ml cyclohexane and 1 ml oleylamine, sonicated for 5 min to accelerate the dissolution, injected into $HAuCl_4$ solution quickly, stirred for 1h at room temperature to obtain 4 nm Au seeds. The Au seeds solution was centrifuged at 12000 rpm for 10 min and then re-dissolved in 4 ml cyclohexane. As the gold atoms are considered to be mobile under electron beam[47,48], a thin layer (2-3 layers) of palladium shell was deposited on the gold particle epitaxially to immobilize the surface atoms. We added 15 ml oleylamine to the above obtained Au seed solution and heated it to 150 °C. 182.9 mg (0.6 mmol) $Pd(acac)_2$ was dissolved in 1 ml oleylamine and quickly injected into the pre-heated Au seed solution, stirred for 2 h, centrifuged at 12000 rpm for 10 min and washed with ethanol 7 times. The final multiply twinned particles was re-dissolved in cyclohexane.

**Data acquisition**

After deposit on thin $Si_3N_4$ film, we performed plasma cleaning on the particles to avoid any possible contamination during the data acquisition. The tilt series of each particle was acquired using an aberration-corrected FEI Titan Themis G2 300 electron microscope with an electron acceleration voltage of 300 keV. Detailed acquisition parameters are listed in the Supplementary Table 1. In order to obtain high quality STEM images and to reduce damage to the sample, a fiducial particle nearby was used to adjust the residual aberration and focus. The electron dose used for the icosahedral nanoparticles was at approximately ~$10^5$ $e^-$/Å$^2$, which has been demonstrated as a safe electron dose to prevent the damage to bimetallic nanoparticles[27, 39–40]. To minimize sample drift, we acquired three sequential images at each angle with a dwell time of 2 μs.

**Image pre-processing and tomographic reconstruction**

The three sequential images acquired at each angle were summed up by cross-correlation to increase SNR and correct sample drift[27]. The images were denoised using the block-matching and 3D filtering (BM3D) algorithm[49]. After denoising, a mask slightly larger than the boundary of the particle was generated with Ostu thresholding. The background intensity within the mask was estimated using Laplace interpolation and then subtracted. The background-subtracted images were aligned in the direction perpendicular to the rotation axis using the center of mass method, and along the rotation axis using the common line method.

The pre-processed images were reconstructed using the real space iterative reconstruction (RESIRE) algorithm[40]. The R factors converged after 200 iterations. Angular refinement and spatial re-alignment were employed to minimize the angular errors due to sample holder rotation and stage instability. After no further angular correction and reconstruction quality improvement, the final reconstructions were computed using the parameters listed in Supplementary Table 1.

**Determination of 3D atomic coordinates and chemical species classification**

The local maxima in the 3D reconstruction were obtained using polynomial fitting[50]. Peak positions were determined by polynomial fitting within 3*3*3 voxels around each local maximum. The possible atomic positions were determined with the constrains of the minimum inter-atomic distance of 3.45 Å. A 3D polynomial fitting is then performed on the possible atomic positions to determine the exact atomic coordinates. The atoms that were unidentified or misidentified due to fitting failure were manually corrected[27]. The manual correction is routinely applied during the atom tracing and refinement in protein crystallography[51].

All atoms are classified into Au and Pd by K-means clustering, based on the integrated intensity within the surrounding 7*7*7 voxels centered at each atom[52]. Due to the effect of missing wedge and noise, some surface atoms were identified as non-atoms due to their weak intensity or the particles were misclassified due to irregular distribution of atomic intensities inside the particles[40]. We performed local re-classification and manual correction of the initially classified model to get the final results[39]. The total number of atoms and the classification of the three particles were shown in the Supplementary Table 1.

**BOO calculation**

The averaged local BOO parameters, such as $Q_4$ and $Q_6$, as well as the normalized BOO parameters for the ideal FCC, HCP models were calculated using the procedure published elsewhere[40,41]. The first-nearest-neighbour shell distance of 3.45 Å was used as a constraint.

**Crystal structure determination**

The number of nearest-neighbor (NN) atoms around each atom, the distance and relative coordinates to the NN atoms are calculated first, and the ideal fcc, hcp, $D_h$ and $I_h$ models containing 12 coordinated atoms are built. We do a dictionary lookup between ideal polyhedra at different orientations, and then the iterative closest point (ICP) algorithm[53,54] is used to search for the best transformations between the coordinated polyhedra of each atom and the ideal model. If the distance between the coordinated atom and its nearest model atom is less than the radius of the atom, the coordinated atom will be paired with the center atom. A similarity score will be given to the center atom based on the similarity between the coordinated polyhedra and the ideal model, where the score for each class of polyhedra is an order parameter defined as

$$s_k = \sum_{j=1}^{12} \max(1 - \frac{|\mathbf{r}_j - \mathbf{r}_k - \mathbf{mp}_j|}{d_{max}}, 0)$$

where $s_k$ is an order parameter for each polyhedra at esch site, $\mathbf{r}_j$ is the position of the $j_{th}$ neighboring coordinate to site k at position $\mathbf{r}_k$, $\mathbf{m}$ is the rotation matrix for the best transformations, $\mathbf{p}_j$ is the polyhedra with 12 vertices, and $d_{max}$ is the maximum allowed distance of a site from an ideal position.

To compare the scores of different crystallographic textures, the similarity score of the center atom is normalized by the total number of NN coordinated atoms. The model with the highest similarity score will be assigned as the crystal type of the center atom. If the highest similarity score of possible crystal type is lower than a score threshold of 0.5, the atom will be classified as undefined. We further checked those undefined atoms based on their local crystallographic texture.

**Solid angle and packing efficiency calculation**

The categorization of atoms into a specific crystal region is dictated by the directional vectors of three edges, originating from the center icosahedral atom in this region. The representative direction for each edge is determined using principal component analysis[55]. After we obtain the directions of three edges represented by the normalized basis $[\mathbf{e_1}\ \mathbf{e_2}\ \mathbf{e_3}]$, the transformed position of every atom relative to the center $I_h$ atom can be written as

$$\mathbf{p}' = [\mathbf{e_1}\ \mathbf{e_2}\ \mathbf{e_3}]^{-1}\mathbf{p}$$

where $\mathbf{p}$ is the position relative to the $I_h$ atom, and $\mathbf{p}'$ is the transformed position. If the elements of $\mathbf{p}'$ are all positive with 1/6 distortion tolerance, the atom is in this current region with basis $[\mathbf{e_1}\ \mathbf{e_2}\ \mathbf{e_3}]$. Two crystal regions shared one common face can be called connected regions. With this connection definition, we can create a graph to represent the connection of 20 crystal regions. The solid angle $\Omega$ of each region can be obtained by the basis vectors as follow

$$\Omega = \frac{|e_1^T \cdot (e_2 \times e_3)|}{|e_1||e_2||e_3| + |e_3|(e_1^T \cdot e_2) + |e_2|(e_1^T \cdot e_3) + |e_1|(e_2^T \cdot e_3)}$$

Packing efficiency of a crystal region can be calculated by the volume occupied by atoms within the alpha shape region. The value of alpha can be set by the first valley of RDF curve. The occupation volume is hard to be calculated directly. As a result, Monte Carlo has been carried out to estimate the packing efficiency. The implementation of the Monte Carlo method is as follows: For each grain within the particle, an encompassing envelope is constructed using the alpha shape method to represent the total volume. Points are randomly and uniformly selected within the space defined by the alpha shape, and the probability of these points falling within the Au atoms is calculated relative to the total number of points. This probability provides the value of packing efficiency.

**Strain calculation**

We employed the similar methods of strain calculation described elsewhere[43]. We employed the Green-Lagrange strain to estimate the strain of this nanoparticle. The strain tensor **E** can be calculated from the elastic deformation gradient **F** as follow

$$\mathbf{E} = \frac{1}{2}(\mathbf{F}^T \mathbf{F} - \mathbf{I})$$

The elastic deformation gradient can be directly obtained from Polyhedron Template Matching (PTM). It should be noticed that only the strain tensor of crystalline part can be obtained according to the principle of infinite strain theory in our system. It is not feasible to find a reference lattice for the disordered part.

**Molecular dynamics (MD) simulation**

Molecular dynamics simulations were performed using the Large-scale Atomic/Molecular Massively Parallel Simulator (LAMMPS). Au nanoparticles, consisting of 3580 atoms were quenched from an initial temperature of 1500 K down to 300 K, beginning in a liquid state. We employed the EAM force field for Au atoms. Each simulation was repeated 100 times to ensure accuracy and reproducibility. From 1500 K to 300 K, we took $1\times10^5$ MD steps for every 100 K dropped, and finally $1\times10^6$ MD steps were taken at 300 K. The final structural configurations are analyzed using the same method as the experimental ICNPs.

# Supplementary Information for

**Janus icosahedral particles: amorphization driven by three-dimensional atomic misfit and edge dislocation compensation**

Zhen Sun[1,3], Yao Zhang[1,3], Zezhou Li[1], Xuanxuan Du[1], Zhiheng Xie[1], Yiheng Dai[1], Colin Ophus[2], Jihan Zhou[1]

[1]*Beijing National Laboratory for Molecular Sciences, Center for Integrated Spectroscopy, College of Chemistry and Molecular Engineering, Peking University; Beijing, 100871, China.*

[2]*National Center for Electron Microscopy, Molecular Foundry, Lawrence Berkeley National Laboratory, Berkeley, CA 94720, USA.*

[3]*These authors contributed equally to this work.*

*Correspondence and requests for materials should be addressed to J. Z. (email:*

*jhzhou@pku.edu.cn)*

This PDF file includes Supplementary Tables 1-7 and Supplementary Figs. 1-16.

Other Supplementary Materials for this manuscript include Supplementary Movies 1-2.

**Supplementary Table 1 | Data collection, processing, reconstruction, refinement and statistic.**

|  | ICNP-1 | ICNP-2 | ICNP-3 |
|---|---|---|---|
| **Data Collection and Processing** | | | |
| Voltage (kV) | 300 | 300 | 300 |
| Convergence semi-angle (mrad) | 30.0 | 30.0 | 30.0 |
| Detector inner angle (mrad) | 39.4 | 39.4 | 39.4 |
| Detector outer angle (mrad) | 200 | 200 | 200 |
| Pixel size (Å) | 0.343 | 0.343 | 0.343 |
| Scanning current (pA) | 30 | 15 | 15 |
| Number of projections | 55 | 56 | 58 |
| Tilt range (°) | -74.0 69.0 | -76.0 76.0 | -73.5 76 |
| Electron dose ($10^5$ e$^-$/Å$^2$) | 5.2 | 2.7 | 2.8 |
| **Reconstruction** | | | |
| Algorithm | RESIRE | | |
| Oversampling ratio | 4 | 4 | 4 |
| Number of iterations | 200 | 200 | 200 |
| **Refinement** | | | |
| R (%)[a] | 0.0765 | 0.0836 | 0.0789 |
| **Statistics** | | | |
| # of atoms | | | |
| Total | 3643 | 4108 | 4711 |
| Au | 2186 | 2145 | 2296 |
| Pd | 1457 | 1963 | 2415 |

[a] The R-factor is defined by $R = \frac{1}{N}\sum_\theta \frac{\sum_{x,y}|\prod_\theta(O)\{x,y\}-b_\theta\{x,y\}|}{\sum_{x,y}|b_\theta\{x,y\}|}$, where $\prod_\theta(O)\{x,y\}$ is the back projection of the reconstruction volume at angle $\theta$, $b_\theta\{x,y\}$ is the real projection image at angle $\theta$, and N is the number of projections.

**Supplementary Table 2 | The atom number of twenty tetrahedra-like domains of ICNP-1**

|          | Domain  | Atom number (serial number) | | | | | | Averaged number |
|----------|---------|----------|----------|----------|----------|----------|----------|------|
| C5 side  | Class 1 | 448 (1)  | 386 (2)  | 459 (3)  | 318 (4)  |          |          | 403  |
|          | Class 2 | 313 (5)  | 263 (6)  | 356 (7)  | 299 (8)  | 281 (9)  | 239 (10) | 292  |
| C5' side | Class 3 | 218 (11) | 200 (12) | 190 (13) | 142 (14) | 155 (15) | 247 (16) | 192  |
|          | Class 4 | 165 (17) | 135 (18) | 167 (19) | 163 (20) |          |          | 158  |

**Supplementary Table 3 | The atom number of twenty tetrahedra-like domains of ICNP-2**

|  | Domain | Atom number (serial number) |  |  |  |  |  | Averaged number |
|---|---|---|---|---|---|---|---|---|
| C5 side | Class 1 | 303 (1) | 334 (2) | 295 (3) | 292 (4) |  |  | 306 |
|  | Class 2 | 348 (5) | 275 (6) | 287 (7) | 269 (8) | 233 (9) | 290 (10) | 284 |
| C5' side | Class 3 | 252 (11) | 267 (12) | 271 (13) | 254 (14) | 284 (15) | 233 (16) | 260 |
|  | Class 4 | 239 (17) | 262 (18) | 263 (19) | 323 (20) |  |  | 272 |

**Supplementary Table 4 | The solid angle of twenty tetrahedra-like domains of ICNP-1**

|  | Domain | Solid angle (serial number) | | | | | | Averaged angle |
|---|---|---|---|---|---|---|---|---|
| C5' side | Class 1 | 50.6° (20) | | | | | | 50.6° |
|  | Class 2 | 39.6° (17) | | 40.5° (18) | | 39.6° (19) | | 39.9° |
|  | Class 3 | 36.4° (11) | 32.7° (12) | 35.2° (13) | 35.3° (14) | 34.7° (15) | 35.8° (16) | 35.0° |
| C5 side | Class 4 | 33.8° (5) | 36.4° (6) | 34.4° (7) | 35.0° (8) | 34.0° (9) | 35.4° (10) | 34.8° |
|  | Class 5 | 33.3° (2) | | 32.3° (3) | | 32.9° (4) | | 32.8° |
|  | Class 6 | 32.4° (1) | | | | | | 32.4° |

**Supplementary Table 5 | The solid angle of twenty tetrahedra-like domains of ICNP-2**

|  | Domain | Solid angle (Serial number) | | | | | | Averaged angle |
|---|---|---|---|---|---|---|---|---|
| C5' side | Class 1 | 45.9° (20) | | | | | | 45.9° |
| | Class 2 | 37.9° (17) | | 42.7° (18) | | 38.0° (19) | | 39.5° |
| | Class 3 | 36.9° (11) | 39.0° (12) | 34.3° (13) | 33.7° (14) | 33.2° (15) | 35.6° (16) | 35.4° |
| C5 side | Class 4 | 35.1° (5) | 35.6° (6) | 33.8° (7) | 33.3° (8) | 34.4° (9) | 36.3° (10) | 34.8° |
| | Class 5 | 34.4° (2) | | 33.1° (3) | | 33.6° (4) | | 33.7° |
| | Class 6 | 33.1° (1) | | | | | | 33.1° |

**Supplementary Table 6 | The packing efficiency of twenty tetrahedra-like domains of ICNP-1**

|          | Domain  | Packing efficiency (serial number) |           |           |           |           |           | Averaged packing efficiency |
|----------|---------|------------|------------|-----------|-----------|-----------|-----------|---------|
| C5 side  | Class 1 | 69.7 (1)   | 72.7 (2)   | 70.5 (3)  | 72.4 (4)  |           |           | 71.3    |
|          | Class 2 | 67.3 (5)   | 72.3 (6)   | 72.0 (7)  | 67.9 (8)  | 70.9 (9)  | 71.4 (10) | 70.3    |
| C5' side | Class 3 | 66.9 (11)  | 71.2 (12)  | 73 (13)   | 69.5 (14) | 72.2 (15) | 71.9 (16) | 70.8    |
|          | Class 4 | 71.2 (17)  | 70.4 (18)  | 63.4 (19) | 62.3 (20) |           |           | 66.8    |

**Supplementary Table 7 | The packing efficiency of twenty tetrahedra-like domains of ICNP-2**

|  | Domain | Packing efficiency (serial number) | | | | | | Averaged packing efficiency |
|---|---|---|---|---|---|---|---|---|
| C5 side | Class 1 | 68.6 (1) | 66.3 (2) | 69.7 (3) | 62.1 (4) | | | 66.7 |
| | Class 2 | 68.7 (5) | 65.3 (6) | 69.9 (7) | 67.4 (8) | 61.8 (9) | 67.9 (10) | 66.8 |
| C5' side | Class 3 | 64.7 (11) | 67.6 (12) | 68.1 (13) | 64.2 (14) | 71.4 (15) | 68.6 (16) | 67.4 |
| | Class 4 | 67.2 (17) | 69.7 (18) | 71.2 (19) | 66.6 (20) | | | 68.7 |

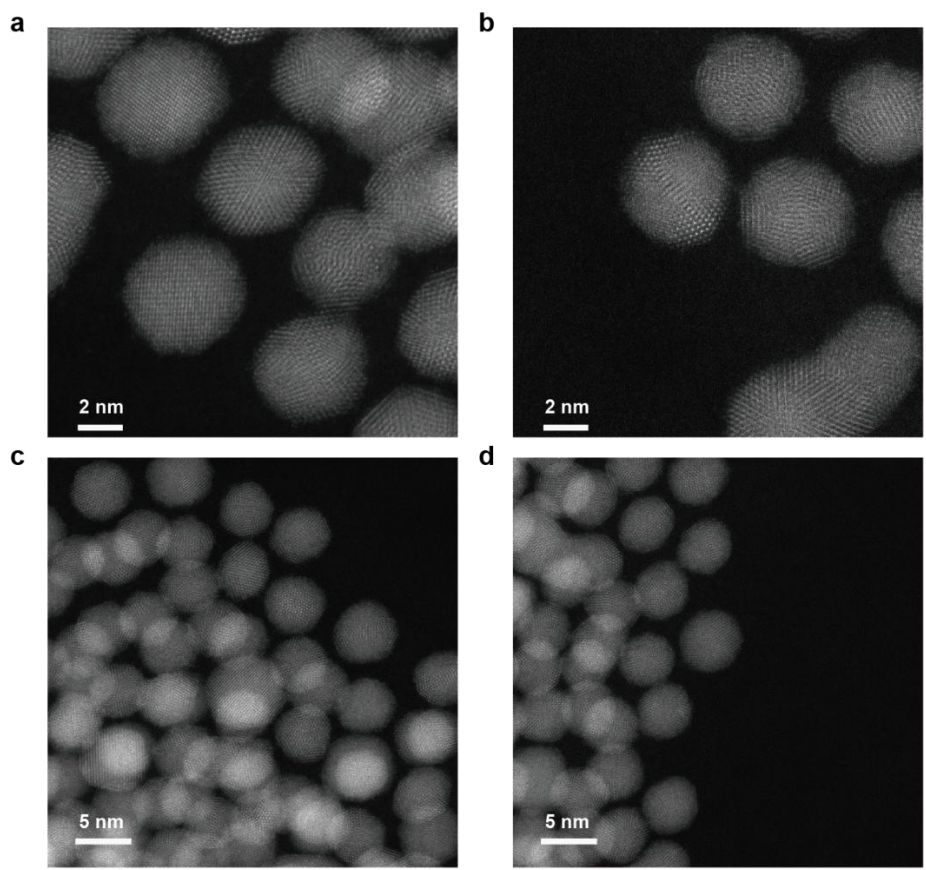

**Supplementary Fig. 1 | The scanning transmission electron microscopy (STEM) images show the multiply twinned particles (MTPs).** MTPs exhibit various features, including fivefold twinning boundaries, stacking faults, and randomly-orientated polycrystalline grains.

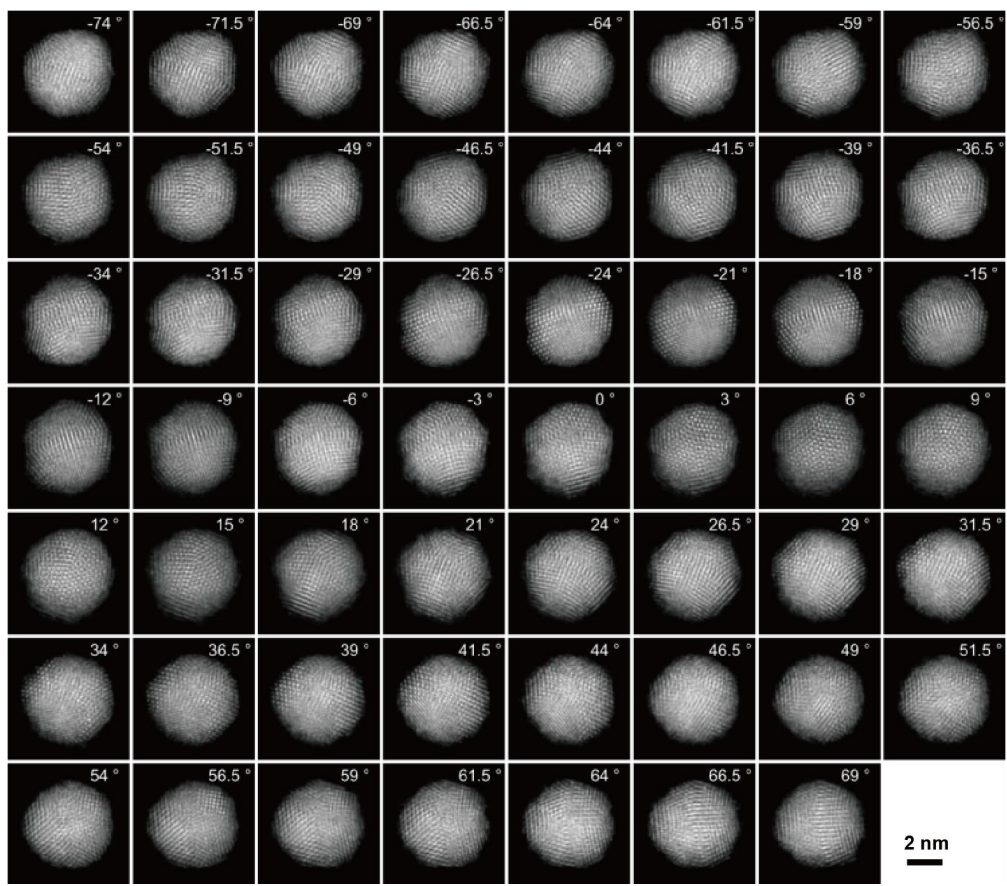

**Supplementary Fig. 2 | Tomographic tilting series of ICNP-1.** 55 ADF-STEM images with a tilt range from −74.0º to +69.0º.

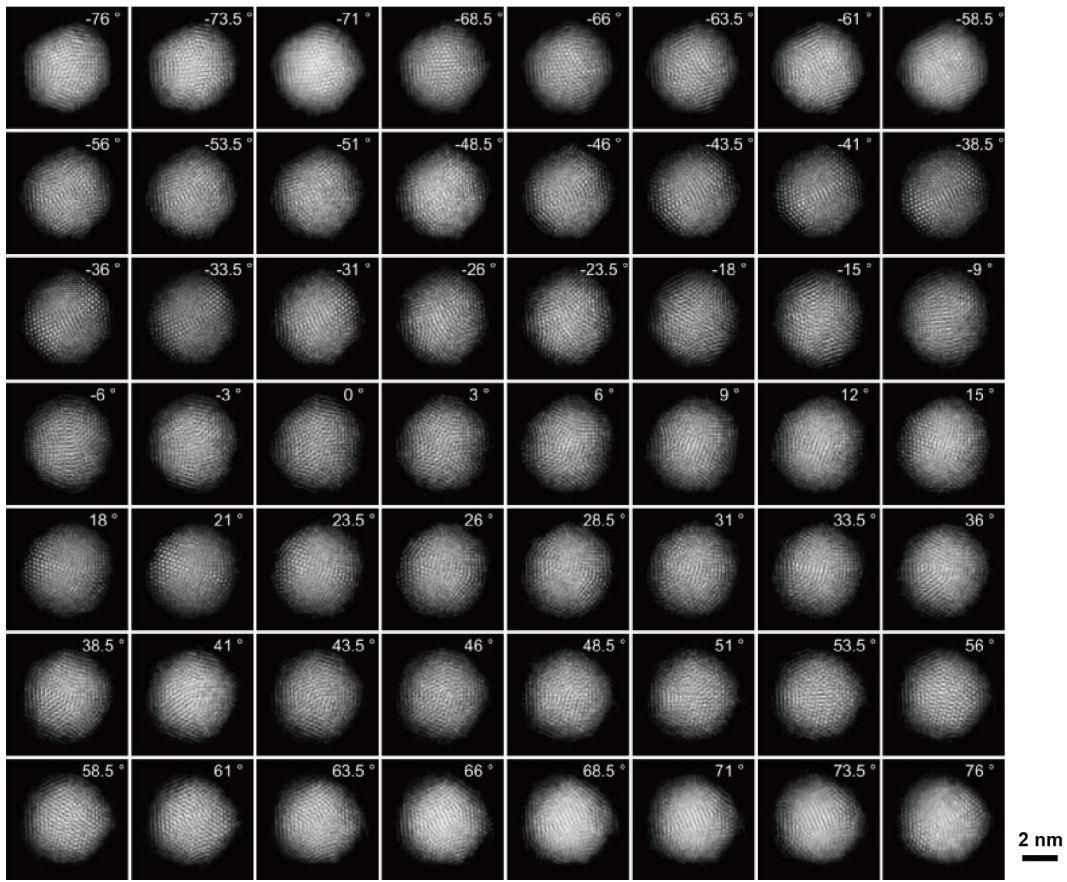

**Supplementary Fig. 3 | Tomographic tilting series of ICNP-2.** 56 ADF-STEM images with a tilt range from −76.0º to +76.0º.

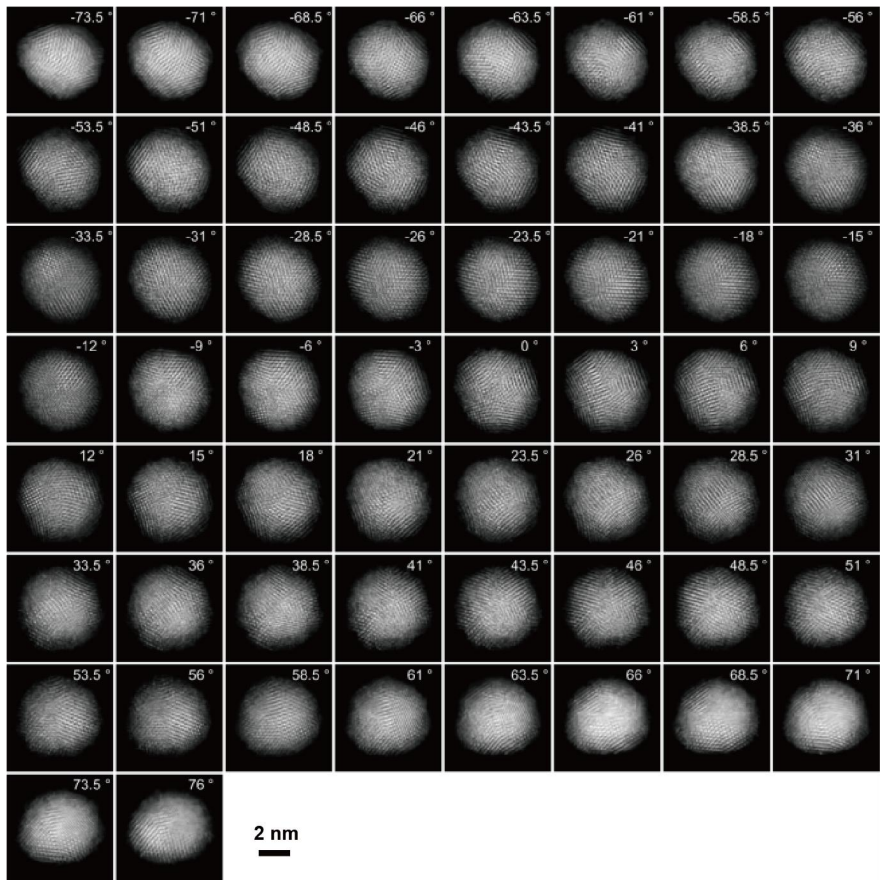

**Supplementary Fig. 4 | Tomographic tilting series of ICNP-3.** 58 ADF-STEM images with a tilt range from −73.5º to +76.0º.

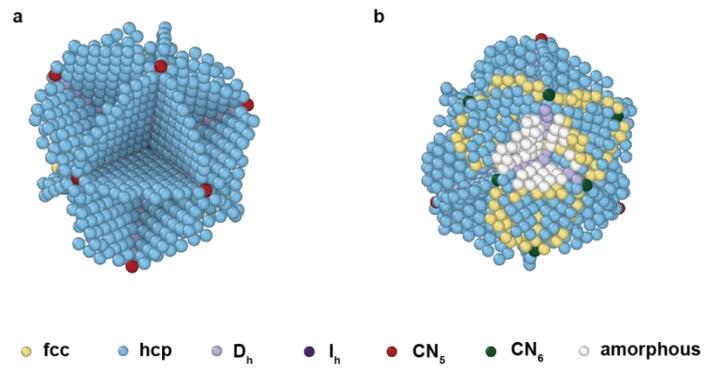

**Supplementary Fig. 5 | 3D atomic structures of the grain boundaries in ICNP-1 viewed from two distinct angles.**

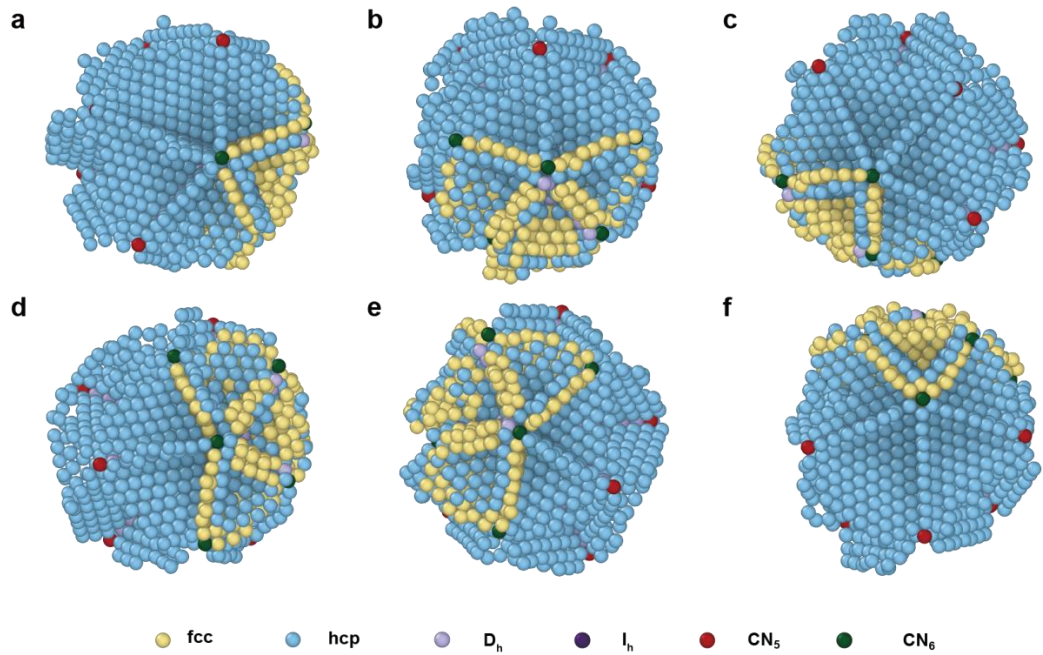

**Supplementary Fig. 6 | The top view of six C5' axes and grain boundaries of ICNP-1.**

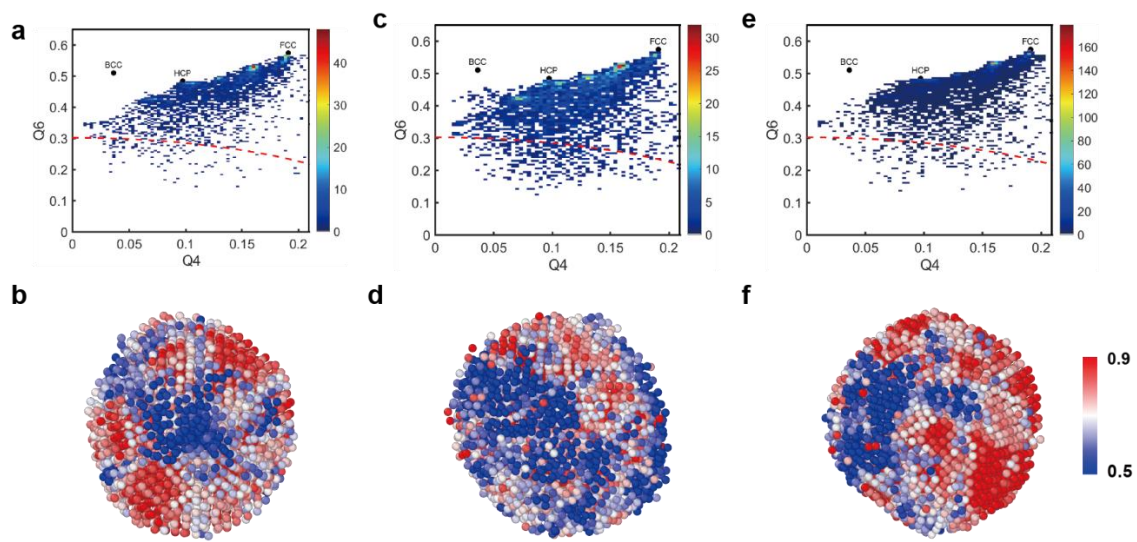

**Supplementary Fig. 7 | Local BOO parameters of all the atoms in ICNP-1 (a), ICNP-2 (c) and ICNP-3 (e).** The distribution of local BOO parameter of all the atoms in ICNP 1, 2 and 3 are shown in (**b**), (**d**) and (**f**), respectively.

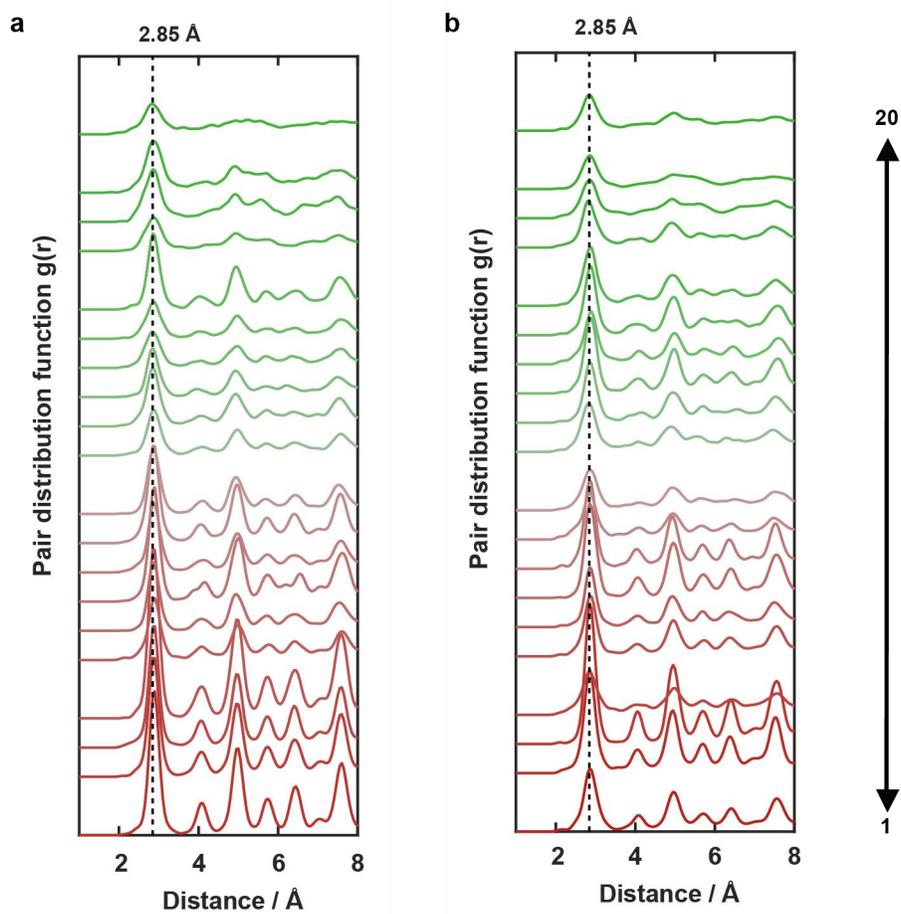

**Supplementary Fig. 8 | Pair distribution functions (PDF) of twenty tetrahedral domains of ICNP-1 (a) and ICNP-2 (b).** The PDF of tetrahedra from C5 side and C5' side are marked in red and green, respectively. The order from bottom to top are the order from tetrahedral domain 1 to 20.

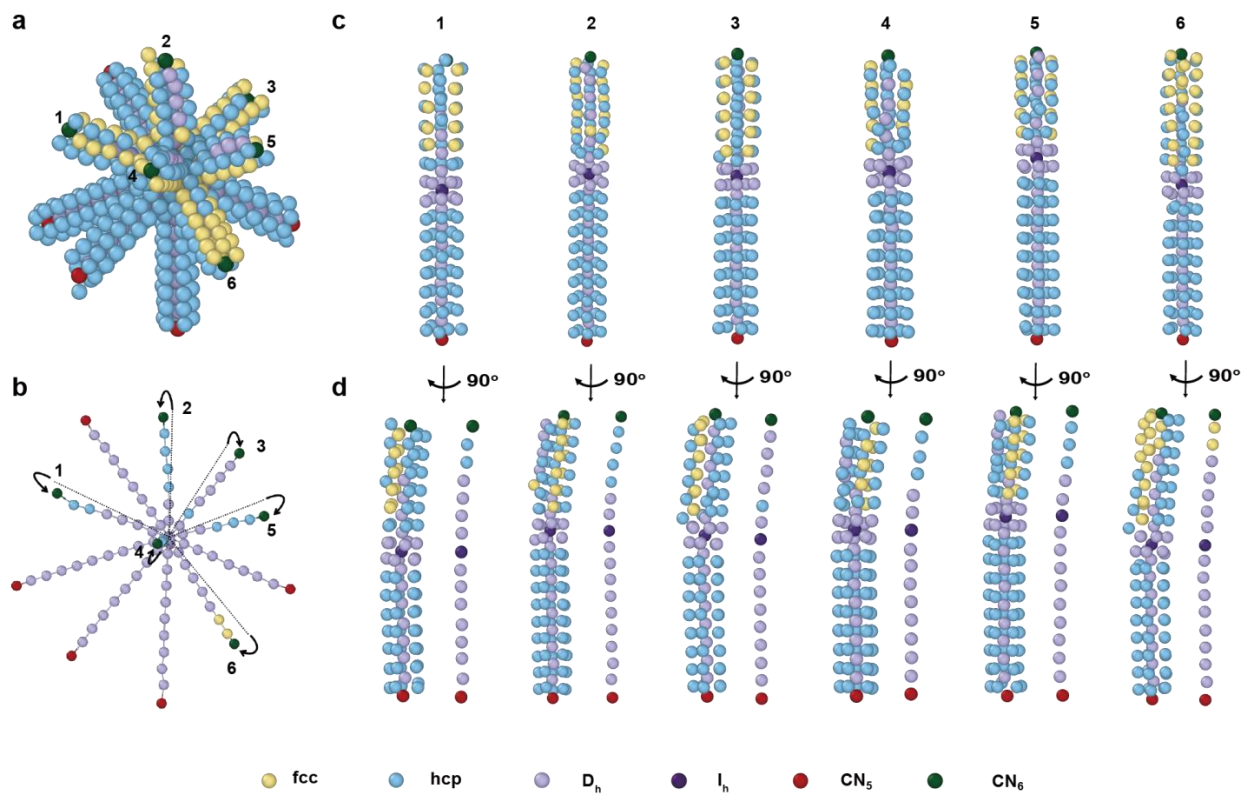

**Supplementary Fig. 9 | The coordination environment of all twelve axes (six C5 axis and six C5' axis) in ICNP-1.** (**a**) All twelve axes and the coordinated atomic columns in ICNP-1. (**b**) The twelve axes (six C5 axes and six C5' axes) in ICNP-1. (**c**) The front view of each group of C5 and C5' axes and the coordinated atomic columns in ICNP-1. (**d**) The side view of each groups of C5 and C5' axes and their coordinated atomic columns. The axial atom columns are shown besides. The ending atoms are colored with red or green intentionally to distinguish C5 and C5' axes.

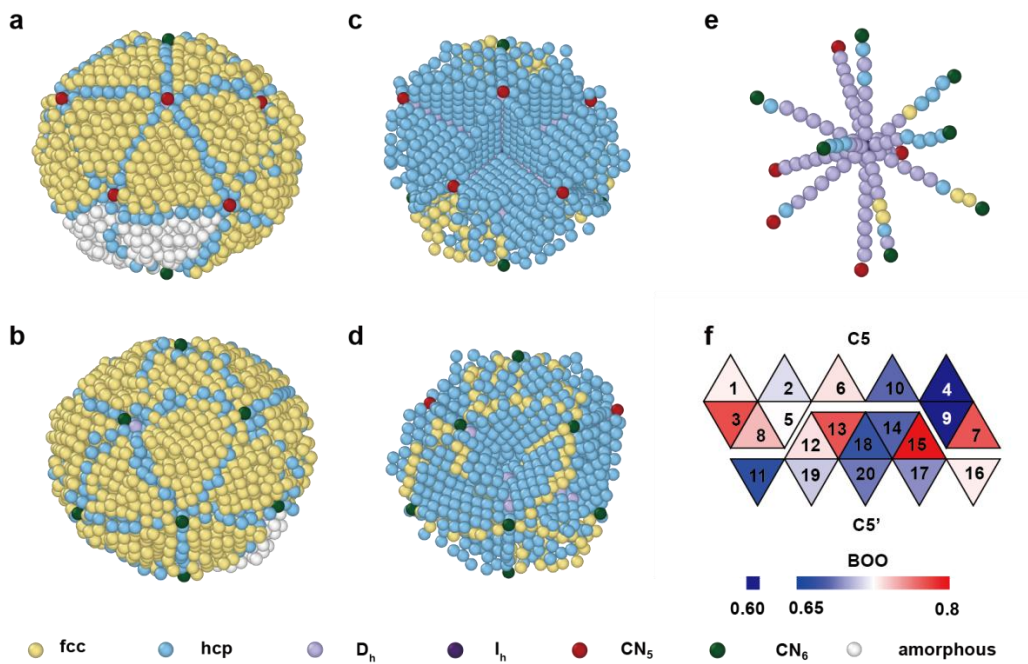

**Supplementary Fig. 10 | 3D atomic structure, fivefold axes and BOO parameter of ICNP-2.** (**a, b**) 3D atomic structure of ICNP-2 viewed from two distinct angles. Two tetrahedral domains at the interface between C5 and C5' sides become amorphous. (**c, d**) The grain boundaries of ICNP-2 viewed from two distinct angles. (**e**) The twelve axes (five C5 and seven C5' axes) in ICNP-2. (**f**) Unwrapped surfaces of twenty domains (each tetrahedral domain has been assigned a number) of ICNP-2 with averaged normalized BOO parameters separated from the C5 and C5' side.

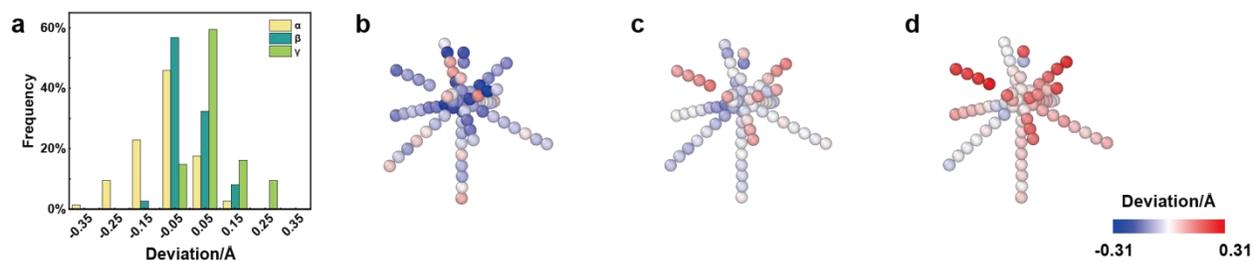

**Supplementary Fig. 11 | The distribution of bond length of α, β, γ in ICNP-1.** (**a**) The global distribution of deviations from the standard Au-Au bond length of α, β, γ in ICNP-1. (**b-d**) The deviations of bond length of α, β, γ, by subtracting the standard Au-Au bond length (2.88 Å), assigned to each $D_h$ type atom in ICNP-1.

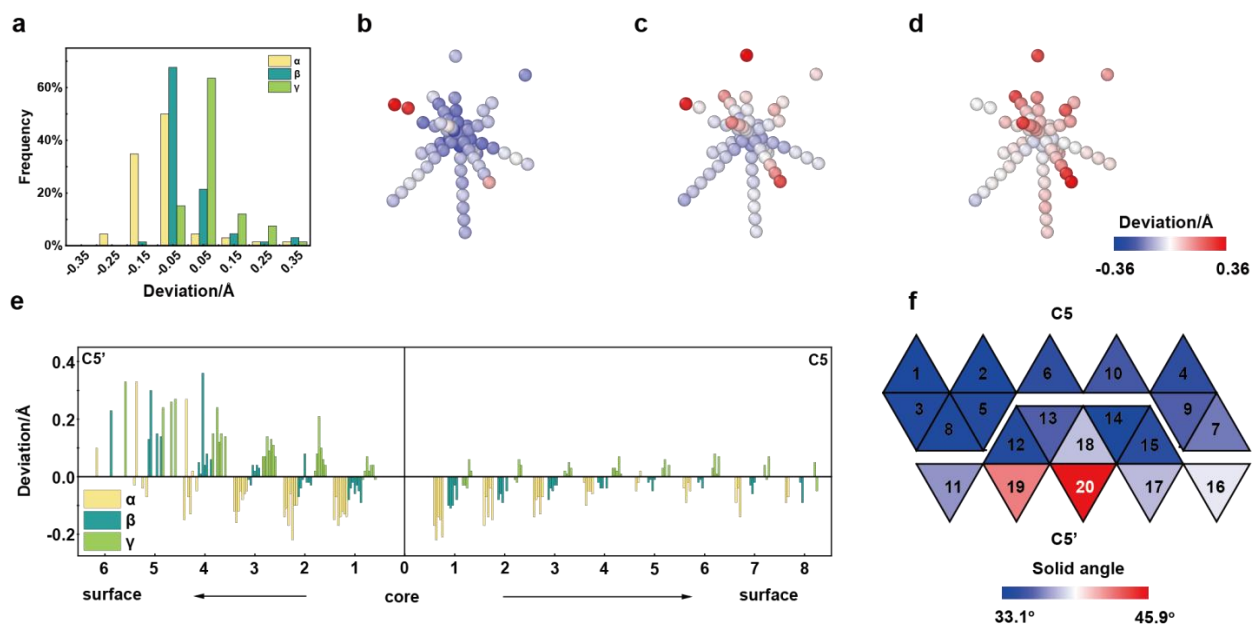

**Supplementary Fig. 12 | The distinct structural characteristics distinguishing the C5 side and the C5' side within ICNP-2.** (**a**) The global distribution of deviations from standard Au-Au bond length of α, β, γ in ICNP-2. (**b-d**) The deviations of bond length of α, β, γ, by subtracting standard Au-Au bond length (2.88 Å), assigned to each $D_h$ type atom in ICNP-2. (**e**) The deviations of bond length of α, β, and γ from the outer surface of C5' axes to the outer surface of C5 axes in ICNP-2. (**f**) Unwrapped surfaces of twenty domains of ICNP-2 with of solid angles in both C5 and C5' sides.

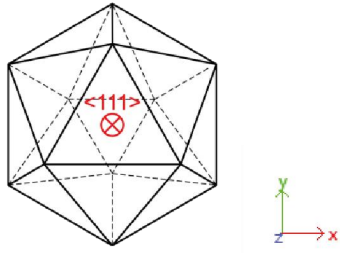

**Supplementary Fig. 13 | Illustration of Z-direction in strain tensor calculation.**

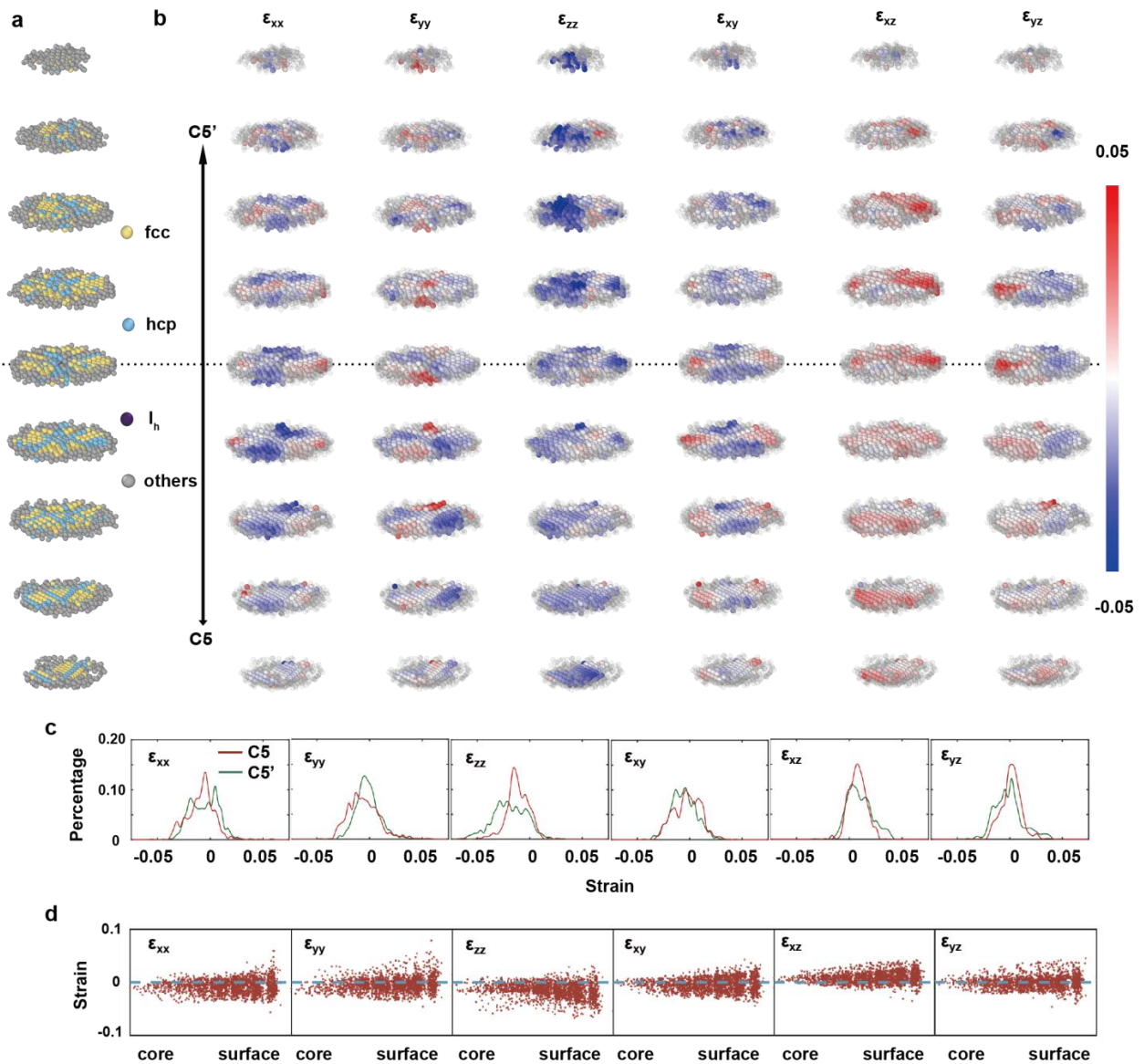

**Supplementary Fig. 14 | The full strain tensor distributions of ICNP-2.** (**a**) Atoms in ICNP-2 used to determine the 3D strain tensor, where the atoms in grey from the amorphous domains and the surfaces are excluded for strain measurement. The Z-axis is determined parallel to the perpendicular direction of the {111} planes shown in Supplementary Fig. 13. (**b**) Maps of the six components of the full strain tensor, with the same block-like distribution as crystal domains in (**a**). The grey atoms show where the strain tensor can not be determined due to the lack of reference lattice. (**c**) The histogram of six components of the full strain tensor on both C5 and C5'

sides. (**d**) Scatter plot of six components of the full strain tensor vs. distance from core to surface, with gradual increase and more scattered distribution.

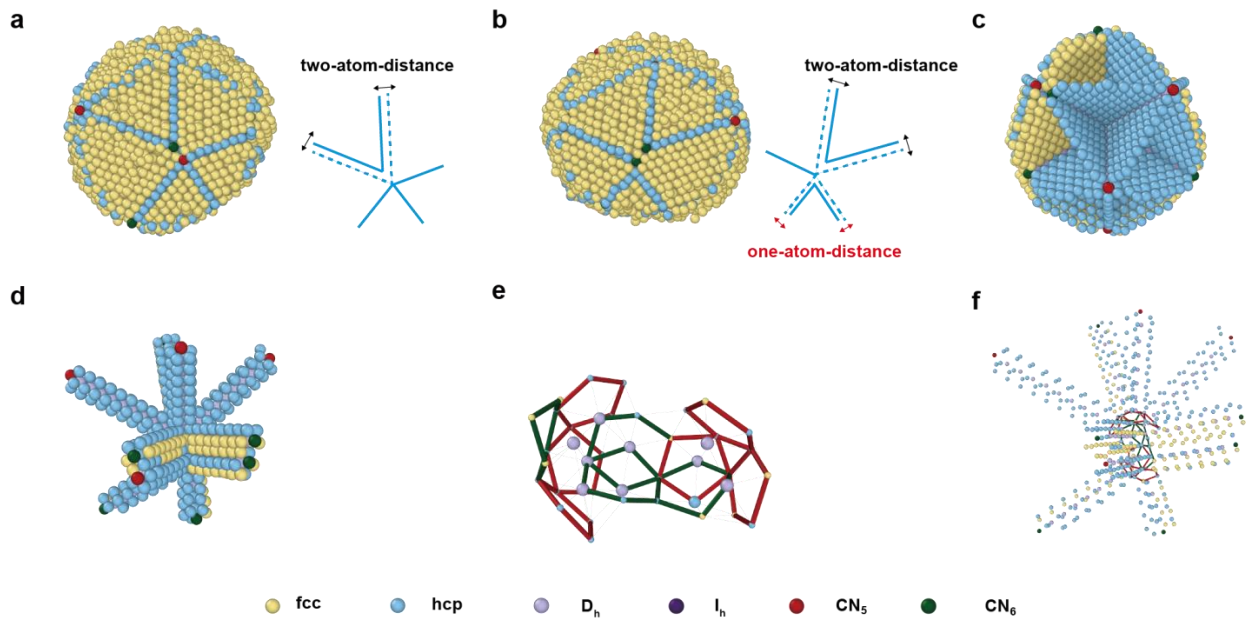

**Supplementary Fig. 15 | 3D atomic structure of ICNP-3.** (**a**, **b**) The morphology of twin axes with C5+C5' (**a**) and C5'+C5' (**b**). Insets depict the slipping distance of the hcp grain boundary. (**c**) The grain boundaries of ICNP-3 viewed from the icosahedra-like side. (**d**) The eight axes, including three fivefold (C5) axes, three pseudo fivefold (C5') axes and two sets of twin axes. (**e**) The disordered boundary formed with fivefold and sixfold skeletons, marked as red and green respectively. (**f**) The combination of eight axes and disordered boundary, showing how the axes end in the a disordered boundary domain.

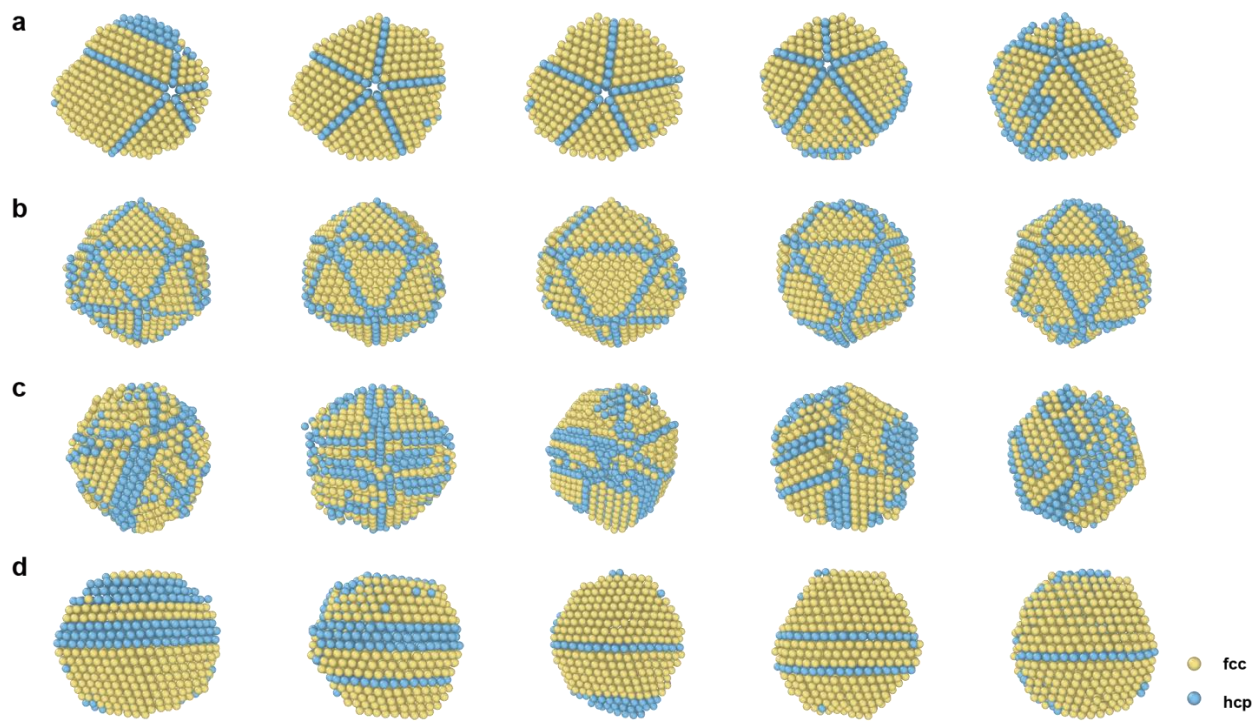

**Supplementary Fig. 16 | The typical four types of annealed gold nanoparticles with different morphologies.** (**a**) Decahedron (DH). (**b**) Icosahedron (IH). (**c**) Polycrystal (PC). (**d**) Crystals with stacking fault (SF). All the particles are displayed with fcc and hcp crystal structure, atoms with other types of structure are not showed.